\documentclass{aa501}
\usepackage[dvips]{graphicx}
\usepackage{natbib}
\bibpunct{(}{)}{;}{a}{}{,}

\newcommand{\HA}{\ensuremath{\mathrm{H}\alpha}}
\newcommand{\HB}{\ensuremath{\mathrm{H}\beta}}
\newcommand{\OII}{\ensuremath{\mathrm{[\ion{O}{ii}]\lambda\mathrm{3727}}}}
\newcommand{\OIII}{\ensuremath{\mathrm{[\ion{O}{iii}]\lambda\lambda
\mathrm{4959,5007}}}}
\newcommand{\NII}{\ensuremath{\mathrm{[\ion{N}{ii}]\lambda\lambda
\mathrm{6548,6584}}}}

\begin{document}

\title{\HA\ emitting galaxies and \\
the star formation rate density  at z$\simeq$ 0.24}
\titlerunning{\HA\ emitting galaxies and the SFRd  at z$\simeq$ 0.24}
\author{S. Pascual \inst{1} \and J. Gallego \inst{1} \and 
A. Arag\'on-Salamanca \inst{2} \and J. Zamorano\inst{1}}
\offprints{S. Pascual}
\mail{spr@astrax.fis.ucm.es}
\institute{Departamento de Astrof\'{\i}sica, Facultad de C.C F\'{\i}sicas,
Universidad Complutense de Madrid, E-28040, Madrid, Spain
\and School of Physics and Astronomy, 
University of Nottingham, University Park, Nottingham NG7 2RD, UK}

\date{Received August 9, 2001; accepted September 20, 2001}

\abstract{
We have carried out a survey searching for H$\alpha$ emitting galaxies
at z$\simeq$0.24 using a narrow band filter tuned with the redshifted
line. The total sky area covered was 0.19 square degrees 
 within the redshift range 0.228 to 0.255 in a set of four fields 
in the ELAIS-N1 zone.
This corresponds to a volume of 9.8$\cdot$10$^3$  Mpc$^3$ and a look-back 
time of 3.6 Gyr when H$_{\mathrm{0}}$=50km s$^{-1}$ Mpc$^{-1}$ and 
q$_{\mathrm{0}}$=0.5 are assumed.
A total of 52 objects are
selected as candidates for a broad band limiting magnitude of $I\sim$ 22.9, 
plus 16 detected only in the narrow band image for a narrow band limiting
magnitude for object detection of 21.0.
The threshold detection corresponds to about 20\AA\ equivalent width
with an uncertainty of $\sim\pm$10\AA. 
Point-like objects (15)  
were excluded from our analysis using
\texttt{CLASS\_STAR} parameter from \texttt{SExtractor}. 
The contamination from other
emission lines such as \OII, \HB\  and \OIII\ at redshifts 
1.2, 0.66 and 0.61 
respectively is estimated, and found to be negligible at the
flux limits of our sample.  We find an 
extinction-corrected H$\alpha$ luminosity density of  
(5.4$\pm$1.1)$\cdot$10$^{39}$ erg s$^{-1}$ Mpc$^{-3}$.
This uncertainty
takes into account the photometric and Poissonian errors only. 
Assuming a constant relation between the H$\alpha$ luminosity and star 
formation rate, the SFR density in the covered volume is  
(0.043$\pm$0.009) M$_{\sun}$ yr$^{-1}$ Mpc$^{-3}$.  This translates to
(0.037$\pm$0.009) M$_{\sun}$ yr$^{-1}$ Mpc$^{-3}$ when the total density is
corrected for the AGN contribution as estimated in the local
Universe. 
This value is a factor $\sim4$ higher than the local SFR density.
This result
needs to be confirmed by future spectroscopic follow-up observations.
}
\maketitle

\section{Introduction}
The star formation rate density of the Universe is one of the key observables
needed for our understanding of galaxy formation and  evolution.  In a key
reference paper, \citet{1996MNRAS.283.1388M} connected the high-redshift
luminosity density obtained from the coaddition  of the emission from
individually-detected galaxies  with that obtained from low redshift surveys.
This luminosity density was then translated into a star formation rate (SFR)
density. Their original SFR density versus redshift plot showed that the SFR
density  steeply increases from $z\sim0$ to  $z\sim1$ and decreased beyond
z$\simeq$2.5, suggesting that it probably peaked between z$\simeq$1 and
z$\simeq$2.  Deep redshift surveys also suggest that the  star-formation
activity substantially increases with redshift until z$\simeq$1
\citep{1994ApJS...94..461S, 1996MNRAS.280..235E, 1996ApJ...460L...1L,
1997ApJ...481...49H, 1998ApJ...504..622H}. However, the high redshift behaviour
is not so clear, and the decline in SFR density beyond $z\sim2$ is still 
contentious.

Detailed theoretical works are starting to shed light to the
problem.  It is now possible to build models which, within the
hierarchical clustering scenario, put together dark matter, gas and
stars \citep[e.g.,][]{1991ApJ...381...14L,1993MNRAS.264..201K,
1994MNRAS.267..981K,1999MNRAS.303..188K,1994MNRAS.271..781C}.  These
models can provide a reasonable match to both the present-day
characteristics of galaxies \citep{1998ApJ...498..504B}, as well as
the properties of galaxies at high redshift
\citep{SP,2001MNRAS.320..504S}.
These models are also able to quantitatively predict the global star
formation history of the Universe, i.e.  the comoving number density
of galaxies as a function of star formation rate, and as a function of
redshift.

One of the major problems that arises when analysing galaxy
populations at different redshifts is how to make a meaningful
comparison.  To test directly whether substantial evolution in the
star-formation activity has occurred, we need to measure the SFR
density of the Universe at different redshifts using similar
techniques. Optimally, we should try to use the same selection
criteria, same galaxy populations and same SFR tracer.  Such a uniform
measurement would provide a much stronger constraint for galaxy
formation and evolution models.

The \HA\ luminosity, related to the number of massive stars, is a
direct measurement of the current star formation rate (modulo the Initial
Mass Function). 
Metallic nebular lines such as \OII\ and \OIII\ (affected by
excitation and metallicity) and far-IR fluxes (affected by dust
abundance and properties) are star-formation
\emph{indicators} rather than \emph{quantitative} measurements
\citep[see, e.g.,][]{1989AJ.....97..700G,1992ApJ...388..310K}.
Thus the best way to quantify current star formation is by using an \HA\
selected sample of galaxies \citep{1998ngst.conf..135C}.
Although star formation in heavily obscured regions
will not be revealed by \HA, if we select the galaxies with the
same criteria at all redshifts, the samples --and the derived
SFRs-- will be directly comparable.

A few pioneering works have estimated average SFR densities measuring
\HA\ luminosities in the near-infrared for small samples below z=1
(\citealt[][ using tunable filters]{2001ApJ...550..593J}), at z$\sim$1
\citep{1999MNRAS.306..843G,1999ApJ...519L..47Y}
and z$\sim$2
\citep{2000PASJ...52...73I,2000A&A...362....9M,2000A&A...362..509V},
and \HB\ luminosities for samples of 5 and 19 objects at z$\sim$3
\citep{1998ApJ...508..539P,astro-ph/0102456}.
These preliminary results indicate that SFRs from Balmer lines follow
 the general trend traced by UV broad-band luminosities at high redshifts.
They are found to be 2--3$\times$ higher than those inferred from the 
extinction corrected UV at all redshifts. 
The UV flux comes from the OB stars on the 
star-forming region and the underlying population whereas 
the \HA\ flux comes from the \ion{H}{ii} region surrounding the OB stars.
The different origin of the radiation explains the different measured
SFRs.

A z$\sim$0 benchmark in this field is the SFR density obtained from the
Universidad Complutense de Madrid survey of H$\alpha$ emission line galaxies in
the local Universe \citep{1995ApJ...455L...1G}.  At z$\sim$0.2, the only
available  \HA\ luminosity function is the one obtained for the  broad-band
selected CFRS sample  \citep{1998ApJ...495..691T}.  In this paper we describe a
survey for H$\alpha$ emitting galaxies at z$\simeq$0.24 carried out with a
narrow band filter.

Section 2 describes the data acquisition and the image reduction;
Section 3 shows the galaxy selection process; in Sect.~4 we calculate 
the \HA\ luminosity function and the SFR density, 
and finally in Sect.~5 the 
conclusions are presented.  All along the paper we assume a Friedman model 
cosmology (cosmological constant $\Lambda_{\mathrm{0}}$=0) with a 
Hubble constant H$_{\mathrm{0}}$= 50 km s$^{-1}$
Mpc$^{-1}$ and deceleration parameter q$_{\mathrm{0}}$=0.5.

\section{Data acquisition and reduction}
Our survey for \HA\ emitting galaxies was carried out using the focal
reducer CAFOS\footnote{http://www.caha.es/CAHA/Instruments/index.html} 
at the 2.2m telescope in CAHA (Centro Astron\'omico Hispano-Alem\'an, 
Almer\'{\i}a, Spain). 

This instrument is equipped with a 2048$\times$2048 Site\#1d CCD with  24$\mu$m
pixels  (0\farcs53 on the sky), which covers a circular  area of 16\arcmin\ 
diameter. In the data  reduction process, the covered area is reduced to
14\farcm6. Four fields were observed, all of them located near the centre of
the European Large-Area ISO Survey field ELAIS-N1 
\citep{1999ESASP.427.1011R,2000MNRAS.316..749O}. Our strategy was to observe 
overlapping regions between the fields, so we can have consistent photometry on
all the frames. All four fields were observed through a 16 nm FWHM narrow band
filter  centred at 816 nm, in a region of low OH emission. Broad band
$I$-filter images were also obtained. The filters used were, respectively,
816/16 and 850/150c in the CAHA filters  database. 

Of the four observing nights, two were lost due to Sahara's dust on 
the atmosphere. The other two nights had  non-photometric conditions. 
The overall seeing was 1\farcs2. Total exposures were 600s in the broad 
and 3000 -- 6000s in the narrow filter. The survey covered 0.19 square 
degrees, corresponding to 9.8$\cdot$10$^3$ Mpc$^3$ 
comoving volume at  z=0.242, given the width of the narrow-band filter.
Table~\ref{tab:fields} shows the fields surveyed.

\begin{table}
\caption{Survey fields}
\begin{tabular}{lllll}
Field Name & RA     & Dec     & \multicolumn{2}{l}{Exposures (s)} \\
           & (J2000)& (J2000) & $I$ &  NB  \\
\hline
\hline
ELAIS a3 &16:05:30  &+54:30:36  & 600 & 3000\\
ELAIS a4 &16:04:00  &+54:30:36  & 600 & 6000\\
ELAIS b3 &16:06:15  &+54:17:36  & 600 & 5000\\
ELAIS b4 &16:04:45  &+54:17:36  & 600 & 4000\\
\hline
\hline
\end{tabular}
\label{tab:fields}
\end{table}

The images were processed using the standard reduction procedures for 
de-biassing and flat-fielding found in the {\tt CCDRED} facility within 
\texttt{IRAF}\footnote{IRAF is distributed by the National Optical Astronomy 
Observatories, which is operated by the Association of Universities for 
Research in Astronomy, Inc.\ (AURA) under cooperative agreement with the 
National Science Foundation.}. Fringing was present in the broad band images
at $\sim$5\% of the sky level. It was removed by combining deep blank sky 
frames to obtain the fringe pattern, placing it at zero mean level, scaling 
it to the level of the sky background of the science frame,  and subtracting. 
The frames were aligned using  the coordinates of bright stars on the images
before combining. Due to the use of the focal reducer, the borders of the 
frames suffer from geometrical distortion and it is not enough to apply a 
shift to align the images. We have applied a general transformation using 
shifts (about 30''), rotation ($\sim$2$\cdot$10$^{-3}$\%) and re-scaling 
($\sim$0.1\%). Finally the frames were combined to obtain the final image for 
each filter used. The frames were scaled to a common count level 
using stars in the overlapping regions between the images.
Because the nights were not photometric, 
photometric calibration was achieved using stars of the USNO-A2 catalogue
\citep{USNO-A2.01} contained in our fields 
and using synthetic colours $B-R$ and $R-I$ calculated with
template spectra from \citet{1998PASP..110..863P}.
We estimate that the zero-point uncertainty is 
$\simeq0.1$ mag.
The zero-point of the narrow band calibration was obtained assuming a 
mean $I-$m$_{NB}$=0, using the bright end of the selection diagrams 
(Figure~\ref{fig:trumpet}). 

\section{Nature of candidates}
\subsection{Object detection and candidate selection}
Catalogues of the objects in all the four surveyed fields were made using 
\texttt{SExtractor} \citep{1996A&AS..117..393B}. The objects are detected
using the \emph{double-image mode}: the narrow band frame is used as a 
reference image for detection and then the flux is summed up in 6 pixel 
diameter apertures in both the narrow- and broad-band image. This
aperture size is 3\farcs18, i.e.,  2.65$\times$FWHM of the seeing, 
corresponding to 15 kpc at z=0.242.

Candidate line emitting objects were selected using their excess
narrow versus broad flux on a plot of m$_{NB}$ versus $I-$m$_{NB}$. For
each candidate, both broad and narrow aperture fluxes, line equivalent width
and the sigma excess are calculated.
The flux density in each filter can be expressed as the sum of the line
flux and the continuum flux density 
(the line is covered by both filters):
\begin{equation}
f^B_{\lambda}=f^c_{\lambda}+\frac{f_L}{\Delta_B}\qquad
f^N_{\lambda}=f^c_{\lambda}+\frac{f_L}{\Delta_N}
\end{equation}
with $f^c_{\lambda}$ the continuum flux; $f_L$ the line flux; $\Delta_B$ and
$\Delta_N$ the broad and narrow band filter effective 
widths and $f^B_{\lambda}$ and 
$f^N_{\lambda}$ the flux density in each filter.
Then the line flux, continuum flux and equivalent width can be express as 
follows:
\begin{eqnarray}
f_L&=&
\Delta_N\left(f^N_{\lambda}-f^B_{\lambda}\right)
\frac{1}{1-\frac{\Delta_N}{\Delta_B}}\\
f^c_{\lambda}&=&
f^B_{\lambda}\frac{1-\frac{f^N_{\lambda}}{f^B_{\lambda}}
\frac{\Delta_N}{\Delta_B}}{1-\frac{\Delta_N}{\Delta_B}}\\
\label{eq:ew}
EW&=&\frac{f_L}{f^c_{\lambda}}=\Delta_N\left(\frac{f^N_{\lambda}-f^B_{\lambda}}
{f^B_{\lambda}}\right)\left(\frac{1}{1-\frac{f^N_{\lambda}}{f^B_{\lambda}}
\frac{\Delta_N}{\Delta_B}}\right)
\end{eqnarray}
The effective widths 
of the filters are calculated as the integral of the transmission
of the filter multiplied by the quantum efficiency of the CCD:
\begin{equation}
\Delta_{filter} = \int T_{filter}\times QE \; d\lambda
\end{equation}
For the broad band filter $\Delta_B$=1665 \AA; 
for the narrow band filter $\Delta_N$=173 \AA.
The conversion flux-magnitude was done using the spectral energy distribution
of Vega given by \citet{1994A&A...281..817C}.

\begin{figure*}
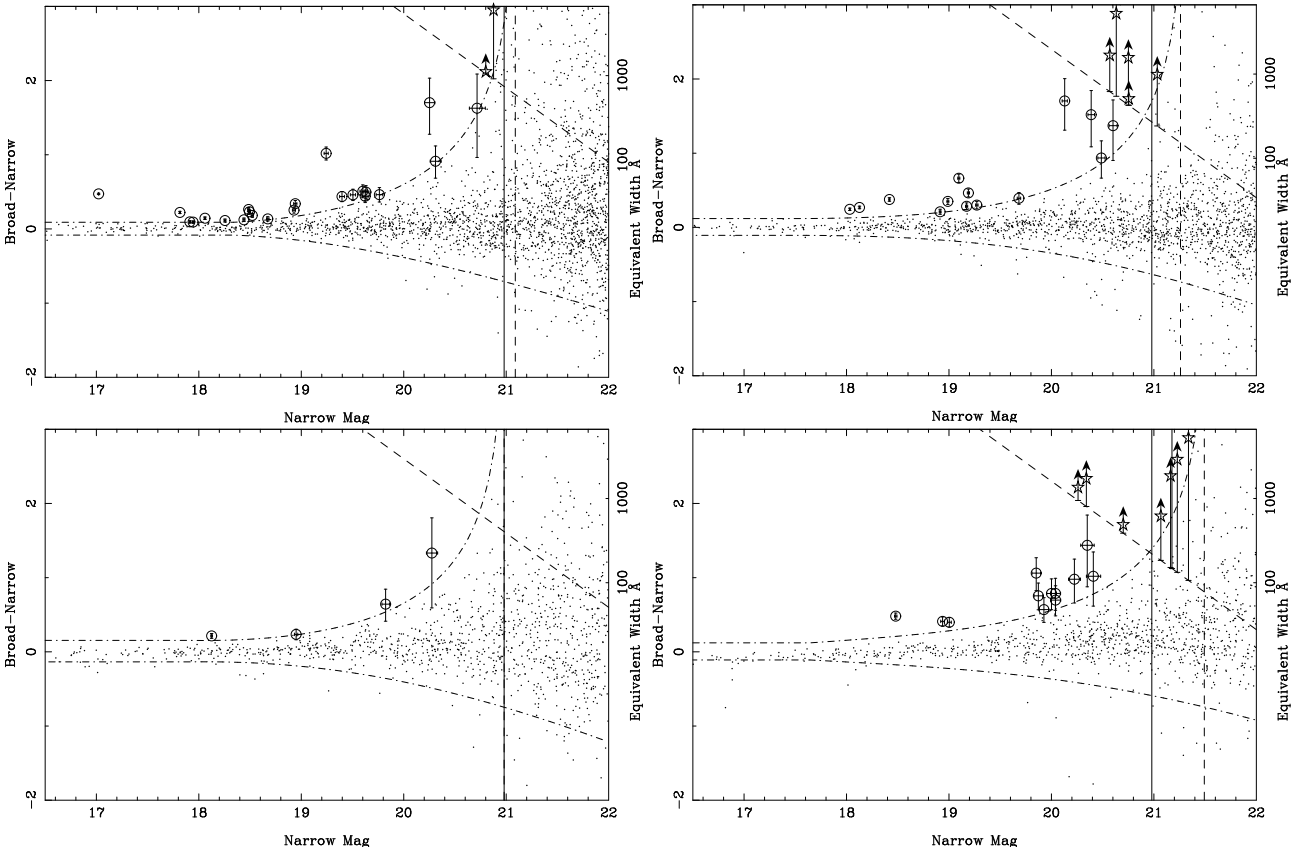

\centering
\includegraphics[angle=-90,width=8.5cm]{1804f1_a3.eps}
\includegraphics[angle=-90,width=8.5cm]{1804f1_a4.eps}
\includegraphics[angle=-90,width=8.5cm]{1804f1_b3.eps}
\includegraphics[angle=-90,width=8.5cm]{1804f1_b4.eps}
\caption{m$_{NB}$ vs $I-$m$_{NB}$ for the fields a3, a4, b3 and b4 (
from top to bottom and left to right). 
The dashed curve represents the 3$\sigma$ detection limit, the diagonal 
dashed line is the limiting broad magnitude, the dashed vertical 
line is the limiting narrow magnitude for object selection, and the solid 
vertical line is the limiting magnitude 
used in the analysis (i.e., the limiting magnitude for the 
shallowest field). 
The star-shaped points with up-pointing arrows 
are objects selected below the limiting broad band magnitude, i.e., detected
in the narrow-band image but only marginally detected in the broad-band image.}
\label{fig:trumpet}
\end{figure*}

In Fig.~\ref{fig:trumpet} we show the plot for the four surveyed 
fields showing the
curve for flux excess 3$\sigma$. Several clear objects exhibiting excess  
emission are shown. The dashed vertical line is the narrow band limiting
magnitude, defined as the m$_{NB}$ that makes
$-2.5\log(1-3\sigma(m)) \to\infty$. 
Under this magnitude no object can be 
selected. The solid vertical line is the brighter of the detection-limiting 
magnitudes on the four fields. Only objects brighter than this limit
will be used to obtain the luminosity function.

Table~\ref{tab:objects} lists the candidates detected in 
the fields at $\geq3\sigma$. The first column (1) is the 
identification number in the catalogue produced by \texttt{SExtractor}; 
(2) and (3) are the coordinates of the object, with an accuracy better
than 1''; (4) and (5) are the magnitudes of the objects inside the apertures,
with typical accuracies better than 0.1 magnitudes for both bands;
(6) is the the equivalent width measured using eq.~\ref{eq:ew}. 
The uncertainty is better than 30\% for low EWs ($<400$\AA).
(7) is the $\sigma$ excess of the detection; and (8), (9) 
are the \texttt{CLASS\_STAR} parameter produced by 
\texttt{SExtractor} in the $I$ and narrow band images.

\begin{table}
\caption{Objects selected in the field ELAIS a3 with measurements}
\tabcolsep=0.03cm
\begin{tabular}{lcccccccc}
& \multicolumn{2}{c}{Coordinates (J2000)} &\multicolumn{2}{c}{Mag} &EW& 
$\sigma$&\multicolumn{2}{c}{\texttt{CLASS\_STAR}}\\ 
ID   & RA  & DEC & $I$ & NB & (\AA) & exc& $I$&NB\\
(1)&(2)&(3)&(4)&(5)&(6)&(7)&(8)&(9)\\
\hline
\hline
  551 & 16:04:41.4 & +54:34:25.52 &  18.7 &   18.5 &     40 &  5.0 & 0.02 & 0.03\\
  561 & 16:04:43.4 & +54:34:35.55 &  21.9 &   20.2 &   1300 &  4.5 & 0.02 & 0.01\\
 1245 & 16:04:52.8 & +54:28:08.54 &  19.8 &   19.4 &    100 &  4.2 & 0.04 & 0.03\\
 1305$^{\mathrm{b}}$ & 16:04:53.9 & +54:28:23.40 &  19.2 &   18.9 &     50 &  4.4 & 0.98 & 0.98\\
 1344$^{\mathrm{b}}$ & 16:04:54.5 & +54:28:16.98 &  20.1 &   19.6 &    100 &  3.4 & 0.98 & 0.98\\
 2247 & 16:05:06.6 & +54:37:59.13 &  20.0 &   19.5 &    110 &  3.9 & 0.02 & 0.03\\
 3420$^{\mathrm{b}}$ & 16:05:17.9 & +54:36:10.93 &  18.0 &   17.9 &     20 &  3.2 & 0.98 & 0.98\\
 4084 & 16:05:23.9 & +54:34:45.12 &  21.2 &   20.3 &    300 &  3.1 & 0.00 & 0.01\\
 5901 & 16:05:35.1 & +54:27:04.54 &  18.8 &   18.5 &     50 &  7.2 & 0.03 & 0.03\\
 6224$^{\mathrm{b}}$ & 16:05:39.8 & +54:26:19.25 &  18.8 &   18.7 &     30 &  3.3 & 0.98 & 0.98\\
 7003$^{\mathrm{a}}$ & 16:05:46.3 & +54:37:12.68 & $>$21.6 &  20.9 &  $>$3000 &  3.3 & 0.00 & 0.44\\
 7224 & 16:05:47.2 & +54:25:57.49 &  20.1 &   19.6 &    120 &  3.8 & 0.25 & 0.04\\
 7227$^{\mathrm{b}}$ & 16:05:46.3 & +54:39:11.74 &  17.5 &   17.0 &    110 &  13.3 & 0.36 & 0.93\\
 7605$^{\mathrm{b}}$ & 16:05:50.0 & +54:38:48.02 &  18.4 &   18.2 &     20 &  3.8 & 0.97 & 0.98\\
 7836 & 16:05:51.4 & +54:32:07.63 &  18.7 &   18.5 &     50 &  6.4 & 0.03 & 0.03\\
 7930 & 16:05:52.4 & +54:33:23.82 &  20.1 &   19.6 &    120 &  3.7 & 0.00 & 0.02\\
 8412 & 16:05:54.5 & +54:38:37.76 &  18.0 &   17.9 &     20 &  3.1 & 0.03 & 0.03\\
 8661 & 16:06:07.9 & +54:35:42.45 &  19.3 &   18.9 &     70 &  5.6 & 0.02 & 0.03\\
 8971 & 16:06:03.2 & +54:30:27.56 &  20.2 &   19.8 &    110 &  3.0 & 0.04 & 0.10\\
 9064$^{\mathrm{a}}$ & 16:06:03.3 & +54:28:11.14 & $>$21.6 &  20.8 &  $>$4000 &  3.2 & 0.12 & 0.02\\
 9668$^{\mathrm{b}}$ & 16:06:20.8 & +54:31:50.12 &  18.6 &   18.4 &     20 &  3.8 & 0.19 & 0.77\\
 9856 & 16:06:20.3 & +54:31:49.59 &  22.3 &   20.7 &   1100 &  3.1 & 0.14 & 0.01\\
 9996$^{\mathrm{b}}$ & 16:06:12.6 & +54:29:42.39 &  18.2 &   18.1 &     30 &  4.7 & 0.48 & 0.92\\
10108$^{\mathrm{b}}$ & 16:06:00.5 & +54:26:52.75 &  18.0 &   17.8 &     50 &  7.0 & 0.97 & 0.97\\
10141 & 16:06:18.1 & +54:30:59.20 &  20.3 &   19.2 &    370 &  9.1 & 0.18 & 0.03\\

\hline
\hline
\end{tabular}
\begin{list}{}{}
\item[$^{\mathrm{a}}$] selected under the broad limit magnitude.
\item[$^{\mathrm{b}}$] catalogued as star.
\end{list}
\end{table}

\begin{table}
\addtocounter{table}{-1}
\tabcolsep=0.03cm
\caption{(cont.) Objects selected in the field ELAIS a4 with measurements}
\begin{tabular}{lcccccccc}
& \multicolumn{2}{c}{Coordinates (J2000)} &\multicolumn{2}{c}{Mag} &EW& 
$\sigma$&\multicolumn{2}{c}{\texttt{CLASS\_STAR}}\\ 
ID   & RA & DEC & $I$ & NB & (\AA) & exc& $I$&NB\\
\hline
\hline
  723$^{\mathrm{a}}$ & 16:03:08.7 & +54:32:15.68 & $>$21.2 &  21.0 &  $>$700 &  3.1 & 0.10 & 0.01\\
 1427$^{\mathrm{a}}$ & 16:03:20.1 & +54:29:39.68 & $>$21.2 &  20.6 &  $>$2000 &  4.6 & 0.99 & 0.78\\
 2126$^{\mathrm{b}}$ & 16:03:28.2 & +54:37:53.50 &  19.1 &   18.9 &     40 &  3.1 & 1.00 & 0.98\\
 2158$^{\mathrm{b}}$ & 16:03:28.5 & +54:27:39.67 &  21.8 &   20.1 &   1300 &  5.7 & 0.96 & 0.90\\
 2606$^{\mathrm{b}}$ & 16:03:32.5 & +54:27:02.20 &  19.3 &   19.0 &     80 &  4.7 & 0.98 & 0.98\\
 2734$^{\mathrm{a}}$ & 16:03:33.6 & +54:27:13.30 & $>$21.2 &  20.8 &  $>$1000 &  3.6 & 0.00 & 0.00\\
 2795 & 16:03:34.0 & +54:26:57.43 &  19.6 &   19.2 &    110 &  5.1 & 0.02 & 0.03\\
 8242$^{\mathrm{b}}$ & 16:04:14.6 & +54:25:56.00 &  19.8 &   19.1 &    180 &  7.2 & 0.82 & 0.94\\
 8655$^{\mathrm{a}}$ & 16:04:19.4 & +54:27:09.68 & $>$21.2 &  20.8 &  $>$1000 &  3.9 & 0.00 & 0.04\\
 8934 & 16:04:22.3 & +54:27:12.42 &  21.9 &   20.4 &    900 &  4.5 & 0.00 & 0.02\\
 8942 & 16:04:22.2 & +54:27:18.81 &  22.0 &   20.6 &    700 &  3.6 & 0.00 & 0.02\\
 9041$^{\mathrm{a}}$ & 16:04:23.6 & +54:27:07.30 & $>$21.2 &  20.6 &  $>$2000 &  4.5 & 0.02 & 0.04\\
 9381$^{\mathrm{b}}$ & 16:04:24.7 & +54:27:03.65 &  19.5 &   19.2 &     60 &  3.4 & 0.98 & 0.96\\
 9807 & 16:04:49.8 & +54:31:51.03 &  18.3 &   18.0 &     50 &  5.4 & 0.03 & 0.03\\
 9826$^{\mathrm{b}}$ & 16:04:43.3 & +54:34:33.76 &  21.4 &   20.5 &    300 &  3.2 & 0.01 & 0.96\\
 9989 & 16:04:37.9 & +54:34:46.66 &  18.4 &   18.1 &     60 &  5.8 & 0.03 & 0.03\\
10732 & 16:04:41.3 & +54:34:23.93 &  18.8 &   18.4 &     80 &  6.9 & 0.03 & 0.03\\
11185 & 16:04:36.6 & +54:36:50.56 &  20.1 &   19.7 &     90 &  3.1 & 0.02 & 0.04\\
11748 & 16:04:32.5 & +54:30:35.14 &  19.6 &   19.3 &     60 &  3.4 & 0.03 & 0.46\\

\hline
\hline
\end{tabular}
\begin{list}{}{}
\item[$^{\mathrm{a}}$] selected under the broad limit magnitude.
\item[$^{\mathrm{b}}$] catalogued as star.
\end{list}
Note: a3\_551 and a4\_10732 are the same object
\label{tab:objects}
\end{table}

\begin{table}
\addtocounter{table}{-1}
\caption{(cont.) Objects selected in the fields ELAIS b3 and b4 with measurements}
\tabcolsep=0.03cm
\begin{tabular}{lcccccccc}
& \multicolumn{2}{c}{Coordinates (J2000)} &\multicolumn{2}{c}{Mag} &EW& 
$\sigma$&\multicolumn{2}{c}{\texttt{CLASS\_STAR}}\\ 
ID   & RA & DEC & $I$ & NB & (\AA) & exc& $I$&NB\\
\hline
\hline
13504 & 16:05:59.0 & +54:13:40.60 &  20.5 &   19.8 &    170 &  3.4 & 0.23 & 0.17\\
14173$^{\mathrm{a}}$ & 16:06:10.2 & +54:23:42.57 & $>$21.3 &  19.5 &  $>$4000 &  9.5 & 0.00 & 0.96\\
14456 & 16:06:13.7 & +54:21:48.86 &  19.2 &   18.9 &     50 &  3.1 & 0.01 & 0.03\\
14658 & 16:06:18.6 & +54:17:10.07 &  21.6 &   20.3 &    600 &  3.7 & 0.00 & 0.00\\
14820 & 16:06:20.9 & +54:25:30.20 &  18.3 &   18.1 &     40 &  4.1 & 0.02 & 0.03\\

\hline
\hline
\end{tabular}

\vspace{1cm}

\begin{tabular}{lcccccccc}
& \multicolumn{2}{c}{Coordinates (J2000)} &\multicolumn{2}{c}{Mag} &EW& 
$\sigma$&\multicolumn{2}{c}{\texttt{CLASS\_STAR}}\\ 
ID   & RA & DEC & $I$ & NB & (\AA) & exc& $I$&NB\\
\hline
\hline
 5083 & 16:03:55.0 & +54:20:32.47 &  20.9 &   19.9 &    400 &  5.1 & 0.13 & 0.01\\
 5353 & 16:03:56.0 & +54:19:54.77 &  20.8 &   20.0 &    230 &  3.8 & 0.01 & 0.01\\
 5420 & 16:03:56.6 & +54:19:56.00 &  20.8 &   20.0 &    230 &  3.8 & 0.09 & 0.01\\
 5465$^{\mathrm{a}}$ & 16:03:57.2 & +54:20:53.35 & $>$22.3 &  20.3 &  $>$3000 &  5.6 & 0.38 & 0.01\\
 5526$^{\mathrm{a}}$ & 16:03:57.8 & +54:18:15.48 & $>$22.3 &  21.2 &  $>$400 &  3.7 & 0.00 & 0.00\\
 6574$^{\mathrm{a}}$ & 16:04:01.8 & +54:22:32.49 & $>$22.3 &  20.3 &  $>$2000 &  5.4 & 0.03 & 0.01\\
 6643 & 16:04:01.2 & +54:22:31.65 &  19.4 &   18.9 &     90 &  4.2 & 0.51 & 0.03\\
 6733 & 16:04:02.6 & +54:21:20.63 &  20.6 &   19.9 &    220 &  4.0 & 0.06 & 0.02\\
 7302$^{\mathrm{a}}$ & 16:04:07.6 & +54:17:37.15 & $>$22.3 &  21.2 &  $>$400 &  3.2 & 0.00 & 0.00\\
 7323 & 16:04:06.5 & +54:23:58.02 &  21.8 &   20.4 &    800 &  4.5 & 0.00 & 0.00\\
 8696$^{\mathrm{a}}$ & 16:04:20.7 & +54:16:08.42 & $>$22.3 &  21.2 &  $>$500 &  3.3 & 0.11 & 0.06\\
 9146$^{\mathrm{a}}$ & 16:04:28.1 & +54:13:40.33 & $>$22.3 &  21.1 &  $>$500 &  3.2 & 0.00 & 0.00\\
 9528 & 16:04:32.2 & +54:22:09.57 &  19.4 &   19.0 &     90 &  4.0 & 0.00 & 0.03\\
11002 & 16:04:51.5 & +54:14:36.91 &  20.5 &   19.9 &    140 &  3.2 & 0.00 & 0.00\\
11594 & 16:04:59.0 & +54:26:33.56 &  20.7 &   20.0 &    190 &  3.5 & 0.00 & 0.01\\
11643 & 16:04:58.8 & +54:17:00.85 &  19.0 &   18.5 &    120 &  6.2 & 0.02 & 0.03\\
11683 & 16:05:00.5 & +54:17:00.68 &  21.2 &   20.2 &    300 &  3.9 & 0.00 & 0.01\\
12527 & 16:05:13.1 & +54:15:39.45 &  21.4 &   20.4 &    400 &  3.6 & 0.14 & 0.06\\
12850$^{\mathrm{a}}$ & 16:05:16.6 & +54:25:15.15 & $>$22.3 &  20.7 &  $>$1000 &  3.9 & 0.03 & 0.01\\
13336$^{\mathrm{a}}$ & 16:05:20.5 & +54:16:18.10 & $>$22.3 &  21.3 &  $>$300 &  3.1 & 0.01 & 0.00\\

\hline
\hline
\end{tabular}
\begin{list}{}{}
\item[$^{\mathrm{a}}$] selected under the broad limit magnitude. 
\item[$^{\mathrm{b}}$] catalogued as star.
\end{list}
\end{table}

\begin{figure*}
\resizebox{\hsize}{!}{
\includegraphics[angle=-90]{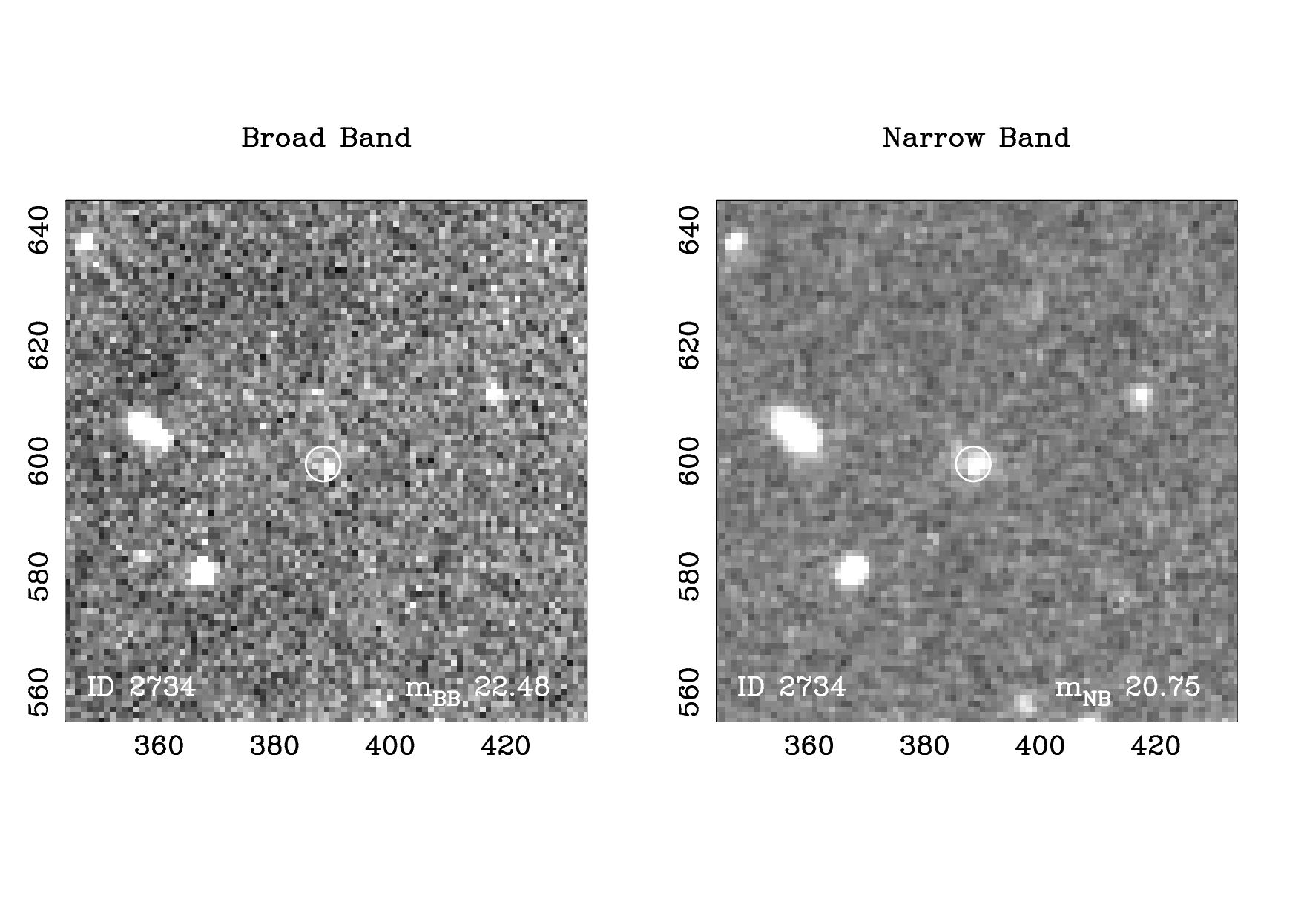}
\includegraphics[angle=-90]{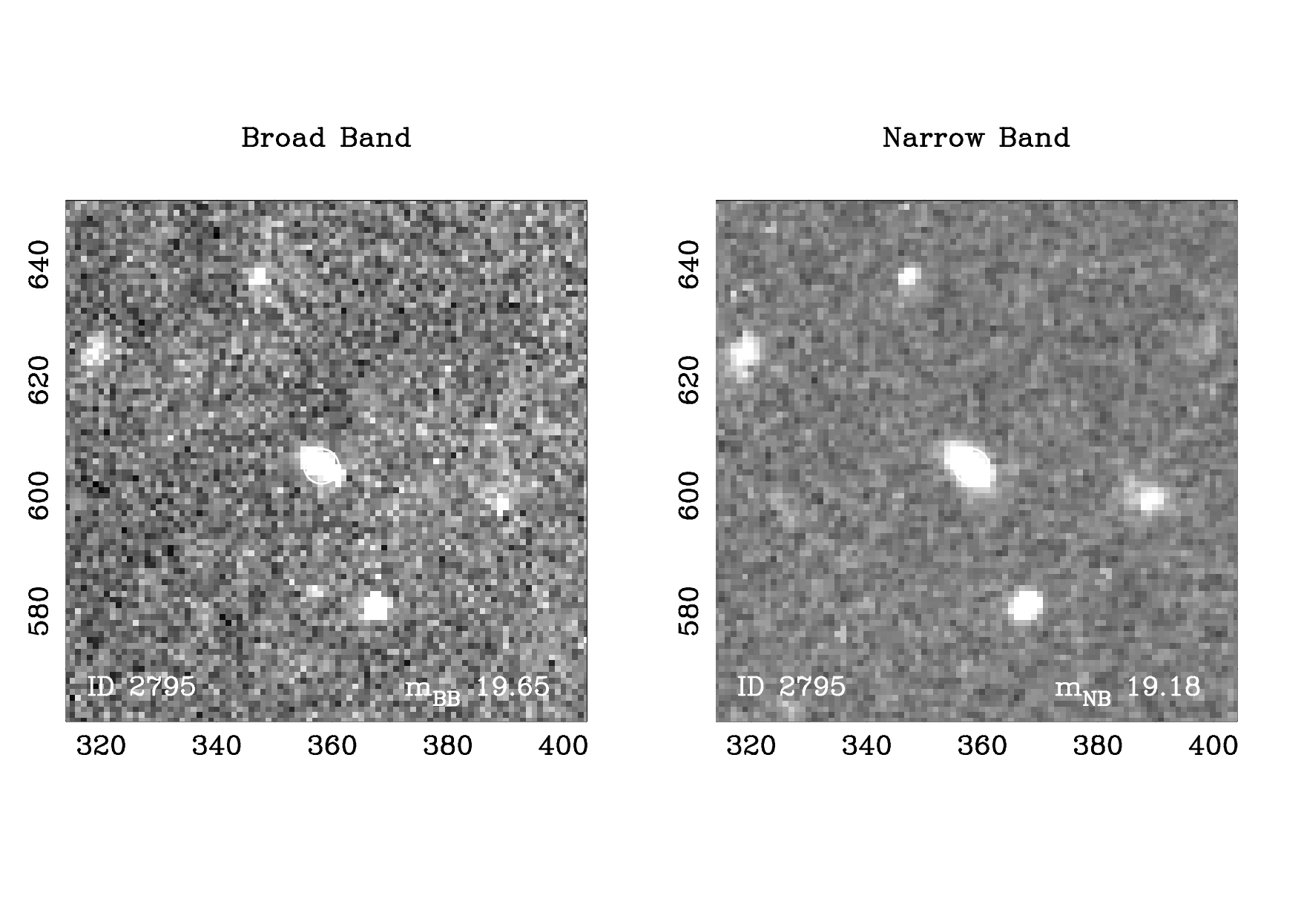}
}

\vspace{-1cm}

\resizebox{\hsize}{!}{
\includegraphics[angle=-90]{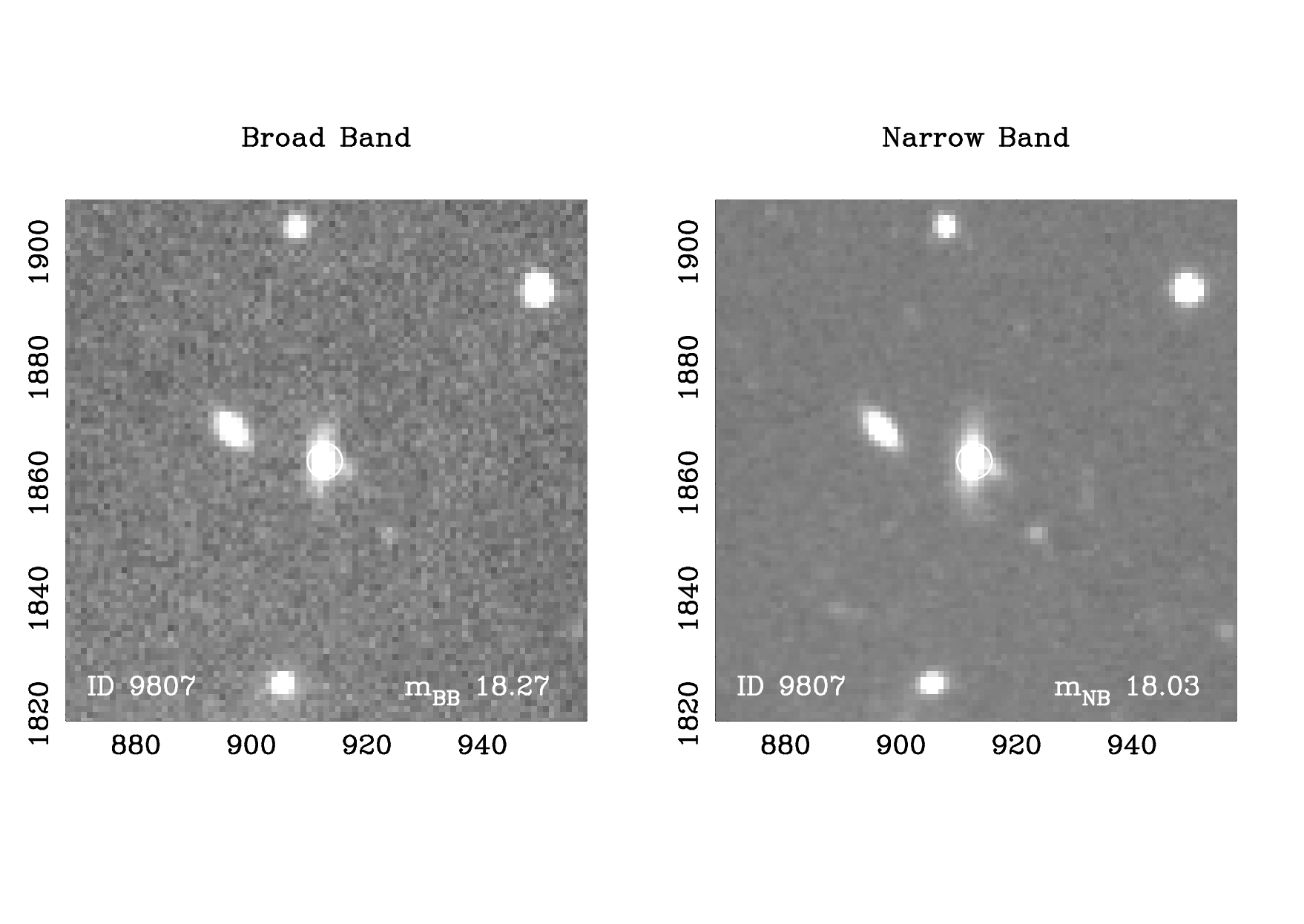}
\includegraphics[angle=-90]{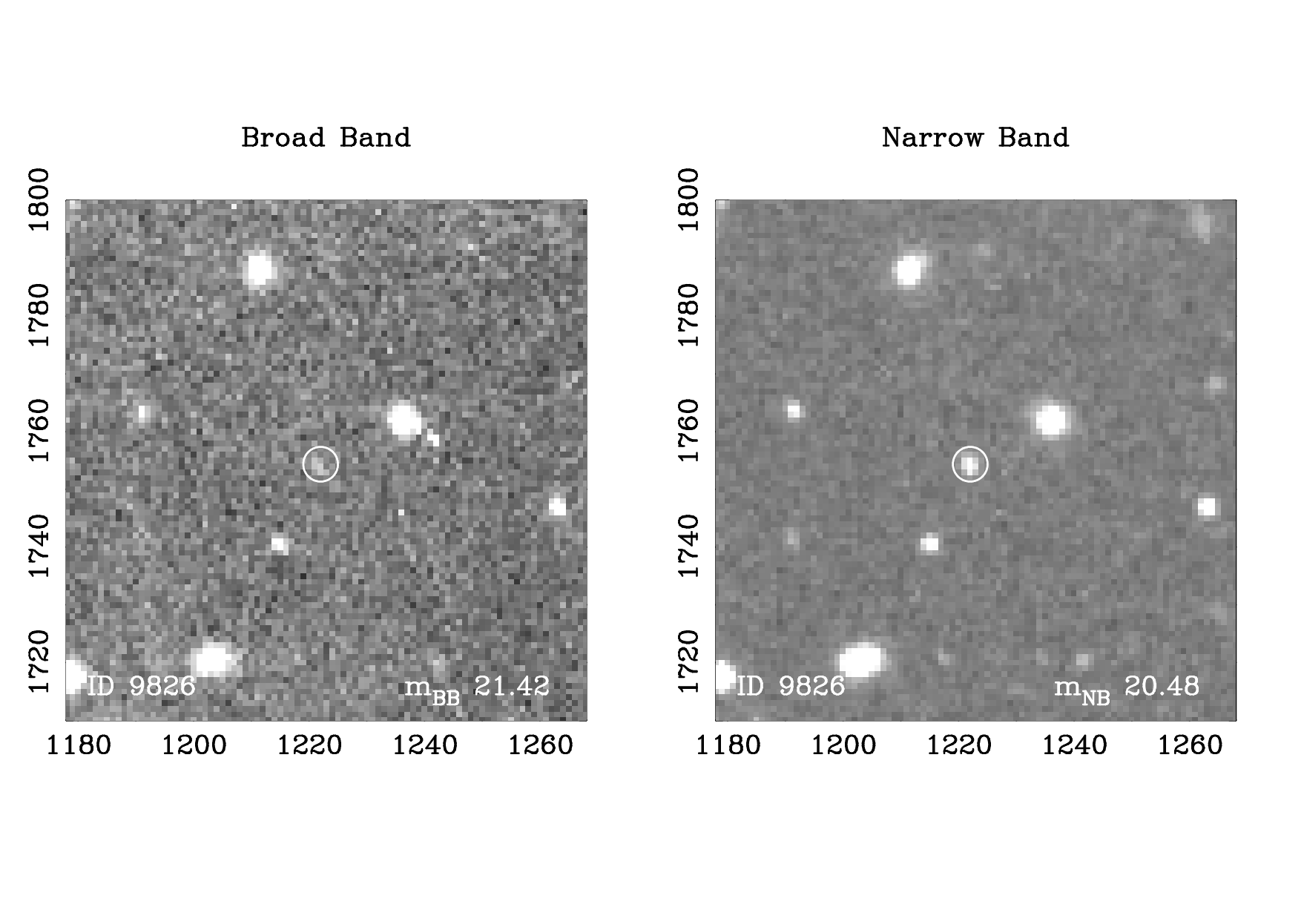}
}
\caption{Sample of selected objects on the field ELAIS-a4. The size of the 
boxes is 25'', north is on the right, east is upwards. Each figure shows
the identification number in the catalogue produced by \texttt{SExtractor}
(ID) and the apparent magnitude in the band}
\label{fig:objects}
\end{figure*}

A total of 52 line-emission candidates were selected in the frames,  with an
additional 16 objects detected only in the narrow band image. The density of
objects selected in both bands is 279 objects per square degree  (365 objects
per square degree when counting also the objects only detected 
in the narrow-band image). 
The objects cover  an $I$ magnitude range from  17 to 22. In 
Table~\ref{tab:objsnum} the range of variation of several quantities of the 
objects detected in each field is listed. In Fig.~\ref{fig:objects} several 
examples of the detected objects are shown. 

\begin{table}
\caption{Number of objects detected in each field}
\tabcolsep=0.12cm
\begin{tabular}{lll@{--}ll@{--}lllll}
Field & \#     & \multicolumn{2}{l}{$I$ mag} & \multicolumn{2}{l}{m$_{NB}$} & 
\multicolumn{2}{l}{EW}& \multicolumn{2}{l}{Limit Mag}\\
ELAIS &  &\multicolumn{2}{l}{range} & 
\multicolumn{2}{l}{range} & \multicolumn{2}{l}{range (\AA)} &$I$ 
& m$_{NB}$\\
\hline
\hline
a3 & 23 & 17.5 & 22.3 & 17.0 &20.7 & 20 & 1300 & 22.9 & 24.2\\
a4 & 14 & 17.2 & 20.8 & 18.0 &20.6 & 40 & 1300 & 22.5 & 24.8\\
b3 & 4  & 18.3 & 21.6 & 18.1 &20.3 & 40 &  600 & 22.6 & 24.3\\
b4 & 12 & 19.0 & 21.8 & 18.5 &20.4 & 90 &  800 & 22.5 & 24.0\\
\hline
\hline
\end{tabular}
Note: Only objects detected in both bands included in the ranges.
\label{tab:objsnum}
\end{table}

\subsection{Galaxy -- star segregation}
\begin{figure}

\resizebox{\hsize}{!}{
\includegraphics{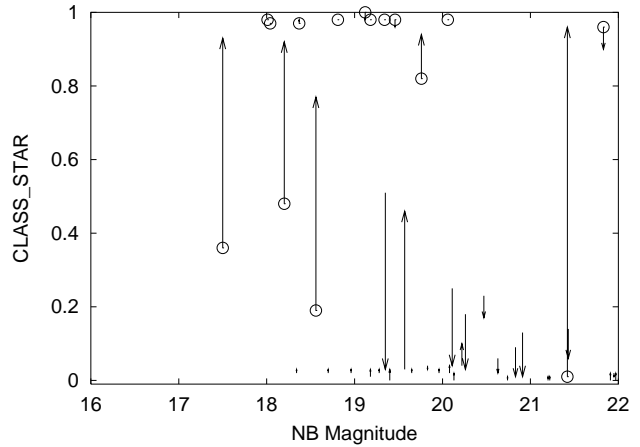}
}
\caption{\texttt{CLASS\_STAR} versus aperture magnitude in the narrow band. 
The tail of the arrow is \texttt{CLASS\_STAR} in the broad band image and
the head is \texttt{CLASS\_STAR} in the narrow band image. The arrows
with a circle in the tail are objects selected as stars.
There is a clear segregation between low \texttt{CLASS\_STAR} 
(non stellar-like profiles) and high \texttt{CLASS\_STAR} (stellar-like 
profiles)} 
\label{fig:stel}
\end{figure}

The sample can be contaminated by stellar
objects (either stars or AGN). To minimise their effect, we have used  the
parameter \texttt{CLASS\_STAR}  provided by \texttt{SExtractor}. We can see in
Fig.~\ref{fig:stel}  how this parameter is distributed. Only the candidates
selected over the limiting magnitudes are plotted, as \texttt{SExtractor} tends
to mis-classify dim objects. Our assumption is that all the objects over 
\texttt{CLASS\_STAR}=0.5 in at least one of the  images are stars (15 objects
in the sample).  The objects with a large variation of the \texttt{CLASS\_STAR}
parameter from the narrow to the broad-band image   were visually
inspected, and classified according to their light profiles. Objects 
classified as stars were excluded from our analysis.

\subsection{Contamination from other lines}
\label{sec:contamin}
A narrow band survey of emission line galaxies can potentially 
detect galaxies with  different emission lines
at different redshifts. If the source redshift and the rest frame wavelength
of the line act to place it inside the narrow band filter, the line
will be detected if it is sufficiently strong. The fixed flux 
detection limit translates  to different luminosities for each line
given the different redshift of the galaxies. Furthermore, 
a different volume is covered for each line for the same reason.
The emission lines we would expect to detect are \HA, \HB, 
\OIII\ and \OII\ \citep{1999MNRAS.310..262T,1992ApJS...79..255K}
as the narrow band filter pass band is too wide to
separate \NII\ from \HA. In Table~\ref{tab:cover} we show
the different redshift coverage for each line.

\begin{table}
\caption{Emission lines potentially detected inside the narrow band}
\tabcolsep=0.12cm
\begin{tabular}{lccccc}

Line & \multicolumn{2}{c}{Redshift range}      & $\bar{z}$& 
d$_c\phantom{1}^{\mathrm{a}}$ & V$\times$ 10$^4\phantom{1}^{\mathrm{b}}$\\
     & \multicolumn{2}{c}{$z_1 \leq z \leq z_2$} & & (Mpc)  & 
(Mpc$^3$)\\
\hline
\hline
\HA\  & 0.228 & 0.256 & 0.242 & 1230  & 0.98\\
\OIII & 0.610 & 0.645 & 0.628 & 2590  & 3.81\\
\HB\  & 0.659 & 0.694 & 0.676 & 2730  & 4.17\\
\OII  & 1.16  &  1.20 & 1.18  & 3870  & 7.34\\
\hline
\hline
\end{tabular}
\label{tab:cover}
\begin{list}{}{}
\item[$^{\mathrm{a}}$] Comoving distance
\item[$^{\mathrm{b}}$] Comoving volume
\end{list}
\end{table}

Since we do not have spectroscopic redshifts for the candidate, it is
necessary to estimate the number of background emission line galaxies
likely to appear in the survey.
We made estimates of the number of galaxies expected
from each of the lines relative to \HA\ \citep{2001ApJ...550..593J}.
For this purpose we need to know the luminosity function in each line and its
evolution with redshift. For \HA, this function has been determined
at a wide range of redshifts \citep{1995ApJ...455L...1G,1998ApJ...495..691T,1999ApJ...519L..47Y}. We also assume an evolution of the 
parameters $\phi^*$, $L^*$ and $\alpha$ of the form:
\begin{eqnarray}
\phi^*(z) & = &\phi^*_0(1+z)^{\gamma_{\phi}}\\
L^*(z)    & = & L^*_0(1+z)^{\gamma_{L}}\\
\alpha(z) & = & \alpha_0+\gamma_{\alpha}z
\end{eqnarray}
using the functional forms adopted by \citet{1997MNRAS.285..613H}. The free 
parameters were constrained using the luminosity function of 
\citet{1995ApJ...455L...1G} at z=0 and \citet{1998ApJ...495..691T} at z=0.2.
The parameters of the \HA\ luminosity function are then:
\begin{eqnarray}
\nonumber
\phi^*(z) & = &10^{-3.2}(1+z)^{4.68}\:\mathrm{Mpc}^{-3}\\
L^*(z)    & = &10^{42.15}(1+z)^{-0.25}\:\mathrm{erg}\:\mathrm{s}^{-1}\\
\nonumber
\alpha(z) & = &-1.3-0.25z
\end{eqnarray}

As the luminosity functions for the other emission lines
have not been determined, our
approach was to scale the L$^*$ parameter of the 
\HA\ luminosity function using mean flux ratios
from \citet{1992ApJ...388..310K}, weighted with the relative occurrence of 
each galaxy type \citep{2001ApJ...550..593J}. The values of $\log L^*$ in 
erg s$^{-1}$ are: 
$\log L^*_{\mathrm{[\ion{O}{iii}]}}$=41.97, $\log L^*_{\HB}$=41.69, 
$\log L^*_{\mathrm{[\ion{O}{ii}]}}$= 42.63.

The rest of the parameters
of the luminosity function are assumed to be equal to the values of the
local \HA\ luminosity function. We have assumed the line ratio 
\HA/\NII=2.3 (obtained by 
\citealt{1992ApJ...388..310K,1997ApJ...475..502G}; 
used by \citealt{1998ApJ...495..691T,1999ApJ...519L..47Y,2000PASJ...52...73I}).

The mean internal extinction is assumed to be
1 magnitude in \HA\ following \citet{1983ApJ...272...54K}, 
who finds this value
for nearby spirals. Also the star-forming galaxies sample from
\citet{1995ApJ...455L...1G} 
has a mean E(B-V)=0.6, which yields a mean \HA\ extinction
of 1 magnitude. The extinction in the another lines is scaled using 
the flux ratios and the extinction law of \citet{1990ARA&A..28...37M}.
These values are 1.78 magnitudes for \OIII, 1.88 for \HB\ and 2.72 for \OII.

\begin{figure}
\resizebox{\hsize}{!}{\includegraphics[angle=-90]{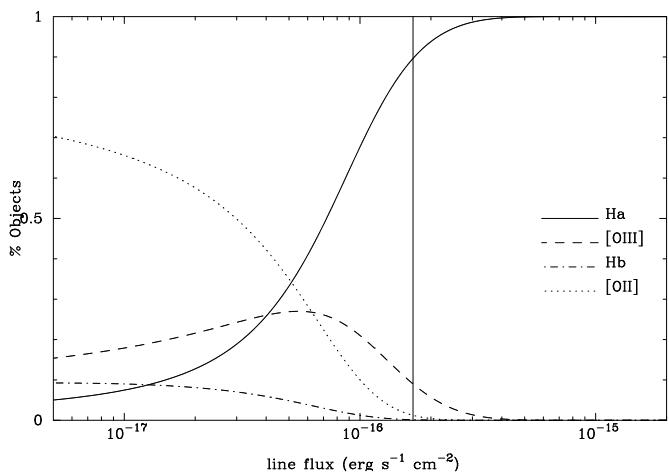}
}
\caption{
\label{fig:counts}
Cumulative number counts for \HA, \HB, \OII\ and \OIII\ emitting galaxies
expected in our sample (see text for details). 
The vertical line shows the line flux in which \HA\ 
galaxies represent more than 90\%
of the total (1.68$\cdot$10$^{-16}$ erg s$^{-1}$)}
\end{figure}

Figure~\ref{fig:counts} shows the predicted cumulative number 
counts for the 
\HA, \HB, \OII\ and \OIII\ as a function of line flux. The population
 of \HA\ emitting galaxies clearly dominates at brighter fluxes as a galaxy 
emitting in other line would 
be more distant, and hence more luminous and rarer. 
The flux line where \HA\ emitters
are more than 90\%
of the total number of galaxies is 1.68$\cdot$10$^{-16}$ 
erg s$^{-1}$ cm$^{-2}$. As the
minimum line flux of the objects in the sample is 
7.22$\cdot$10$^{-16}$ erg s$^{-1}$ 
(corrected from extinction, and  
the effect of the \NII\ lines), 
we can be confident that the contamination
from other emission lines is negligible. 

\subsection{Luminosity of the objects}
\begin{figure}
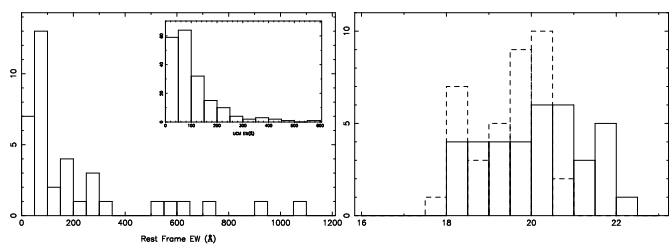

\resizebox{\hsize}{!}{
\includegraphics[angle=-90]{1804f5a.eps}
\includegraphics[angle=-90]{1804f5b.eps}
}

\vskip -3cm
\hskip 2cm
\resizebox{0.25\hsize}{!}{
\includegraphics[angle=-90]{1804f5c.eps}
}
\vskip 1.5cm
\caption{\textbf{ a)} \emph{(left panel)} Histogram of the rest frame EWs 
of our objects. The histogram of the EWs
for UCM survey galaxies is shown in the inset. \textbf{ b)} \emph{(right panel)} 
Histogram of the broad band and 
narrow  band magnitudes of the objects selected in both bands. 
The $I$ magnitudes are represented by solid lines, m$_{NB}$ by dashed lines}
\label{fig:ew}
\end{figure}

At this stage we assume that 
we have a sample of \HA\ emitting galaxies. The distribution of 
equivalent widths and magnitudes of these objects is shown in 
Fig.~\ref{fig:ew}.

The luminosity of the objects is calculated from their line flux. 
We correct for the 
presence of the \NII\ lines, as the narrow filter is unable to separate
the contribution of these lines. We also apply a  mean internal 
extinction correction to the objects. 
For the first two corrections we have assumed the same values used
in Sect.~\ref{sec:contamin}, i.e.,  \HA/\NII=2.33 and A$_{H\alpha}$=1.
(Note that a wide range of \HA/\NII\ ratios is present in star-forming
galaxies. In the extreme case of BCD galaxies, \HA/\NII$\sim$20. 
If the entire sample of detected objects were BCD galaxies, 
a overall increase of the luminosity density of a factor $\sim$1.4 
would occur).
We also apply a small statistical correction (8\%) 
to the measured flux due to the fact that the filter is not square in shape. 
Due the small apparent size of the objects, it is not necessary to make
an aperture correction.
The corrected \HA\ flux is given by:
\begin{equation}
f_0(\HA)=f(\HA)\frac{\HA}{\HA+\NII}10^{0.4A_{H\alpha}}\times 1.08
\end{equation}
Finally the \HA\ luminosity is given by:
\begin{equation}
L(\HA)=4 \pi d^2_l(z) f_0(\HA)
\end{equation}
 using the redshift of the line at the centre of the filter z=0.242 and
$d_l$ the luminosity distance, defined from the comoving distance as 
$d_l=(1+z)d_c$.

\section{Luminosity function and star formation rates}
\subsection{Galaxy luminosity function}
Direct information of the amount and distribution of the 
SFR can be obtained by constructing the luminosity function for galaxies
with star formation activity.
With  all the objects in a small range of redshifts (Table~\ref{tab:cover}),
the luminosity function will be given by:
\begin{eqnarray}
\Phi(\log L_i)=\frac{1}{\Delta \log L}\sum_j \frac{1}{V(z)_j}\\
\nonumber
\mathrm{with} \quad \arrowvert \log L_j-\log L_i\arrowvert<\Delta \log L
\end{eqnarray}
where $V(z)_j$ is the volume of the narrow slice in redshift covered by the 
filter. 
We have taken into account the filter shape in the computation of the volume. 
The correction can be as large as 25\%  
for the faintest galaxies (as compared to a square filter).

\begin{figure}
\resizebox{\hsize}{!}{
\includegraphics[angle=-90]{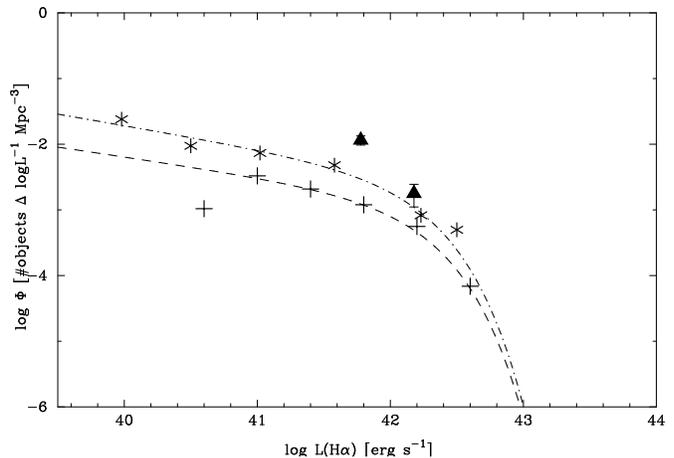}
}
\caption{Luminosity function of the selected objects 
(filled triangles), 
compared with the luminosity functions and Schechter best fit of 
\citet{1995ApJ...455L...1G} (LF crosses, Schechter fit dashed line) and
\citet{1998ApJ...495..691T} (LF asterisks, Schechter fit dot-dashed line).
Error bars indicate Poisson uncertainties only}
\label{fig:lumfunc}
\end{figure}

The summation is over all the galaxies in the \HA\ luminosity range 
$\log L(\HA)\pm\Delta \log L(\HA)$. We have used $\Delta \log L(\HA)$=0.4
(i.e., one magnitude).
Figure~\ref{fig:lumfunc} shows the luminosity function, compared with
the luminosity functions of \citet{1995ApJ...455L...1G} and 
\citet{1998ApJ...495..691T}.

\subsection{Luminosity density and star formation rate density}

The \HA\ luminosity density can be obtained integrating the luminosity 
function:
\begin{equation}
\mathcal{L}=\int^{\infty}_0 \phi^* L \left(\frac{L}{L^*}\right)^
{\alpha}\exp\left(-\frac{L}{L^*}\right)\textrm{d}\!\left(\frac{L}{L^*}\right)
\end{equation}

Given the small luminosity range covered by our objects,  we can not fit a
Schechter function. We  assume a shape ($\alpha$ and $L^*$) for the luminosity
function,  and  obtain the total luminosity density from the summed luminosity
density of the  objects in the sample, extrapolating using the assumed LF
outside the observed range. 
We use the \HA\ luminosity 
function at z=0 obtained by \citet{1995ApJ...455L...1G} in this exercise. 
The line flux of the faintest object 
(7.22$\cdot$10$^{-16}$ erg s$^{-1}$ cm$^{-2}$) translates
into a \HA\ luminosity of 3.78$\cdot$10$^{41}$ erg s$^{-1}$. 
The surveyed volume 
is  9.8$\cdot$10$^3$ Mpc$^3$. 
The summed luminosity density ($\mathcal{L}_{sum}$) in our sample
is (3.3$\pm$0.7)$\cdot$10$^{39}$ erg s$^{-1}$ Mpc$^{-3}$. 
Thus, the total luminosity density can be written as:
\begin{equation}
\mathcal{L}=\mathcal{L}_{sum}\frac{\int^{\infty}_0 L 
\left(\frac{L}{L^*}\right)^{\alpha}\exp\left(-\frac{L}{L^*}\right)
\textrm{d}\!\left(\frac{L}{L^*}\right)}{\int^{\infty}_{L_{lim}} L 
\left(\frac{L}{L^*}\right)^{\alpha} 
\exp{\left(-\frac{L}{L^*}\right)}\textrm{d}\!\left(\frac{L}{L^*}\right)}
\end{equation}
Using the incomplete gamma function
\footnote{defined as 
$\gamma(x,\alpha)=\frac{1}{\Gamma(\alpha)}\int^x_0 u^{\alpha-1}e^{-u}\textrm{d}u$}
finally we obtain:
\begin{equation}
\mathcal{L}=\mathcal{L}_{sum}\frac{1}{1-\gamma(x,2+\alpha)} \quad 
\mathrm{with}\:x=\frac{L_{lim}}{L^*}=0.27
\end{equation}
The total luminosity density is then (5.4$\pm$1.1)$\cdot$10$^{39}$ 
erg s$^{-1}$ Mpc$^{-3}$. Note that using the luminosity function
of  \citet{1998ApJ...495..691T} produces a luminosity density of
(5.9$\pm$1.2)$\cdot$10$^{39}$ erg s$^{-1}$ Mpc$^{-3}$. 
The difference is small ($\sim$10\%) and inside the error bars.
In principle, not all the \HA\ luminosity is produced by star formation. 
The Active Galactic Nuclei can also contribute to the luminosity. 
The amount of this contribution is 8\% of the
number of galaxies and 15\% of the luminosity density for the UCM sample
of \HA\ emitting galaxies. Assuming no evolution in the contribution 
to H$\alpha$ from AGN,  the luminosity density corrected 
from the AGN contribution is 
(4.7$\pm$0.9)$\cdot$10$^{39}$ erg s$^{-1}$ Mpc$^{-3}$.

\begin{figure}
\resizebox{\hsize}{!}{
\includegraphics{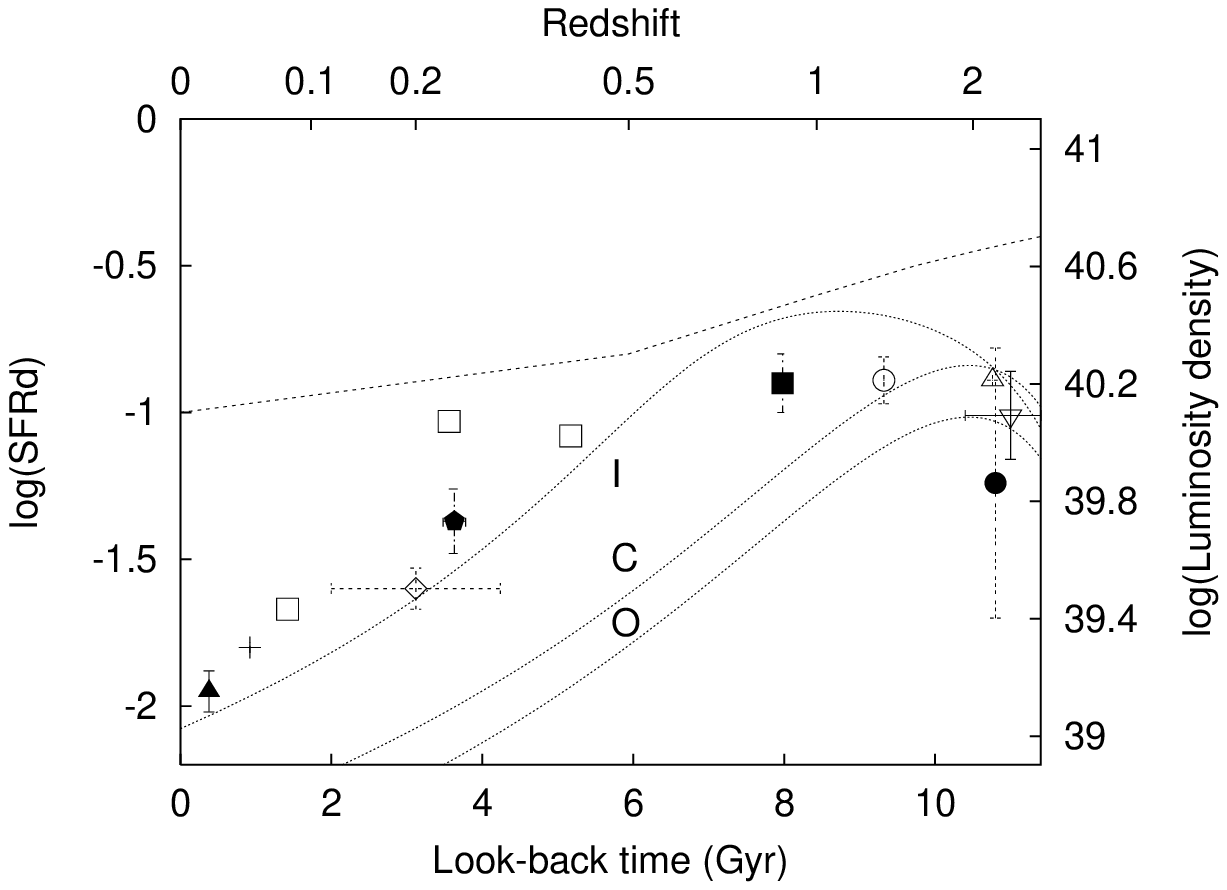}
}
\caption{Evolution of the SFR density measured with Balmer recombination lines.
\HA\ points are from
\citet{1995ApJ...455L...1G} (filled triangle),
\citet{1999adaw.conf..335G} (cross),
\citet{1998ApJ...495..691T} (diamond), 
this work (filled pentagon),
\citet{2001ApJ...550..593J} (open squares),
\citet{1999MNRAS.306..843G} (filled square),
\citet{1999ApJ...519L..47Y} (open circle),
\citet{2000PASJ...52...73I} (open triangle),
\citet{2000A&A...362....9M} (filled circle);
\HB\ point is 
\citet{1998ApJ...508..539P} (inverted triangle).
The top dotted line corresponds to the model 
of \citet{SP}; the rest of the dotted lines are
the inflow (I), closed-box (C) and outflow (O) models 
from \citet{1995ApJ...454...69P}
}
\label{fig:madau}
\end{figure}

The star formation rate can be estimated from the H$\alpha$ luminosity 
using \citep{1998ARA&A..36..189K}:
\begin{equation}
SFR_{\HA}(M_{\sun} \mathrm{yr}^{-1})= 
7.9\cdot10^{-42}L(\HA)(\mathrm{erg}\; \mathrm{s}^{-1})
\end{equation}
Thus, the \HA\ luminosity density translates into a SFR density of 
(0.043$\pm$0.009) 
M$_{\sun}$yr$^{-1}$ Mpc$^{-3}$ (with AGN correction 
(0.037$\pm$0.009) M$_{\sun}$yr$^{-1}$ Mpc$^{-3}$). 
Figure~\ref{fig:madau} shows the evolution of the SFR
density of the Universe from z=0 to z=2.2 measured using Balmer lines.
The right axis shows the luminosity density and the left axis the SFR. 
All the points have been computed using the 
same SFR-luminosity conversion factor and the point from 
\citet{1995ApJ...455L...1G} was computed \emph{with} the AGN contribution.

The \HA\ luminosity is sensitive only to star formation in stars over 
10$M_\odot$, which are the main contributors to the ionising flux.
The SFR density given here is thus 
a extrapolation assuming a given IMF.
The conversion factor between \HA\ luminosity density and SFR density  
is thus very sensitive 
to the IMF, metallicity and details of the population synthesis models 
used \citep{1999MNRAS.306..843G}. It is thus very important to be consistent
when comparing different SFR density estimates from the literature. 
In Fig.~\ref{fig:madau} we have taken the measured H$\alpha$ luminosity
densities and transformed them into SFR densities using the same 
transformation given above. 

The SFR density measured here is lower than that of 
\citet{2001ApJ...550..593J}. At their lower flux cut-off at z$\sim$0.2 
(0.5$\cdot$10$^{-16}$ erg s$^{-1}$ Mpc$^{-3}$) there is a
noticeable contamination from \OIII\ and \HB\ (see Fig~\ref{fig:counts}).
The technique used (tunable filters) implies a very low minimum detected EW 
($\sim$5\AA). Consequently, more objects are detected (either low EW \HA, 
not detected in our survey or
\OIII\ and \HB\ emitting galaxies clasified as \HA). This could explain the 
very high SFR found by \citet{2001ApJ...550..593J} compared to our result.
Another important result at z$\sim$0.2 is that given by 
\citet{1998ApJ...495..691T}. They measured spectroscopic 
\HA+\NII\ fluxes of the $I$-selected CFRS galaxies 
lying at redshift below 0.3. Because of the $I$ selection, these
galaxies have been selected by light from the old stellar population, rather
than from their young stars. This sample is less sensitive to galaxies
undergoing recent star formation than are \HA\ (or $B$) selected surveys. 
Also, the width of the $I$ filter does not favor the detection of glaxies with 
strong \HA\ emission. Despite these differences, our results are consistent
with the strong increase in the SFR density from $z=0$ to $z=1$.

Several theoretical models
are also plotted in Fig~\ref{fig:madau}. 
Semianalytic models of galaxy formation
and evolution \citep{SP}
have difficulties in predicting such steep evolution, while
other approaches like that of \citet{1995ApJ...454...69P} have more success.
It is beyond the scope of this paper to discuss these theoretical models in
any detail since our main concern is to present the observational results. 

\section{Summary and conclusions}

We have carried out a survey searching for H$\alpha$ emitting galaxies at
z$\simeq$0.24 using a narrow band filter tuned with the redshifted line. The
total sky area covered was 0.19 square degrees  within the redshift range 0.228
to 0.255,  corresponding to a volume of 9.8$\cdot$10$^3$  Mpc$^3$ and a
look-back  time of 3.6 Gyr (H$_{\mathrm{0}}$=50km s$^{-1}$ Mpc$^{-1}$ and 
q$_{\mathrm{0}}$=0.5).

A total of 52 objects were  selected as candidates for a broad-band limiting
magnitude of $I\sim$ 22.9,  plus 16 objects detected only in the narrow-band 
image for a narrow band limiting magnitude for object detection of 21.0.  
The threshold
detection corresponds to about 20\AA\ equivalent width with an uncertainty of
$\sim\pm$10\AA.  After excluding  point-like objects  from our analysis, a
sample of 47 emission line galaxies  was produced, 37 of which were detected in
both the narrow-band and broad-band filters. The   minimum line flux in the
sample is  7.22$\cdot$10$^{-16}$ erg s$^{-1}$ cm$^{-2}$,  corresponding to a 
minimum \HA\ luminosity of 3.8$\cdot$10$^{41}$ erg s$^{-1}$. 

In the absence of spectroscopic confirmation,  we have estimated the likely
contamination from other emission lines such as \OII, \HB\  and \OIII\ at
redshifts  1.2, 0.66 and 0.61  respectively, and found it to be negligible at
the relatively high flux limits of our sample.  

We find an  extinction-corrected H$\alpha$ luminosity density of  
(5.4$\pm$1.1)$\cdot$10$^{39}$ erg s$^{-1}$ Mpc$^{-3}$. This uncertainty takes
into account the photometric and Poissonian errors only.  Assuming a constant
relation between the H$\alpha$ luminosity and star  formation rate, the SFR
density in the covered volume is   (0.043$\pm$0.009) M$_{\sun}$ yr$^{-1}$
Mpc$^{-3}$.  This translates to (0.037$\pm$0.009) M$_{\sun}$ yr$^{-1}$
Mpc$^{-3}$ when the total density is corrected for the AGN contribution as
estimated in the local Universe.  This value is a factor $\sim4$ higher than
the local SFR density, and consistent with  the strong  increase in the SFR
density from $z=0$ to $z=1$ previously reported, although   our  results will
have to be confirmed by future  spectroscopic follow-up observations.

\begin{acknowledgements}
This paper is based on observations obtained at the
German-Spanish Astronomical Centre, Calar Alto, Spain,
operated by the Max-Planck Institute fur Astronomie (MPIE),
Heidelberg, jointly with the Spanish Commission for
Astronomy.
This research was supported by the Spanish \emph{Programa Nacional
de Astronom\'{\i}a y Astrof\'{\i}sica} under grant AYA2000-1790. 
S.\ Pascual acknowledges the receipt of a \emph{Formaci\'on de
Profesorado Universitario} fellowship from 
the Universidad Complutense de Madrid. A.\ Arag\'on-Salamanca 
acknowledges generous financial support form the Royal Society. 
This work has benefitted from fruitful discussions with 
C.\ E. Garc\'{\i}a-Dab\'o and P.\ G.\ P\'erez-Gonz\'alez. 
\end{acknowledgements}

\bibliographystyle{apj}
\bibliography{referencias}

\end{document}